\newcommand{\be}{\begin{equation}}
\newcommand{\ee}{\end{equation}}

\documentclass{article}
\usepackage{graphicx}
\usepackage{amsmath}
\usepackage{amsfonts}
\usepackage{amssymb}

\begin{document}

\title{\bf STANDSTILL ELECTRIC CHARGE GENERATES MAGNETOSTATIC FIELD UNDER BORN-INFELD ELECTRODYNAMICS}
\author{S.O. Vellozo$^{1,3}$ \thanks{E-mail: vellozo@cbpf.br}$\;$,  Jos\'{e} A. Helay\"{e}l-Neto$^{1,2}$ \thanks{E-mail: helayel@cbpf.br}$\;$,A.W. Smith$^{1}$ \thanks{E-mail: awsmith@cbpf.br} \\ and L. P. G. De Assis$^{4,2}$ \thanks{E-mail: lpgassis@ufrrj.br}\\
$^{1}${\it  \normalsize Centro Brasileiro de Pesquisas F\'{\i}sicas -- CBPF,} \\
{\it \normalsize Rua Dr.\ Xavier Sigaud 150, 22290-180, Rio de Janeiro, RJ, Brasil}  \\
$^{2}${\it \normalsize Grupo de F\'{\i}sica Te\'{o}rica Jos\'e Leite Lopes,} \\
{\it \normalsize P.O.\ Box 91933, 25685-970, Petr\'opolis, RJ, Brasil} \\
$^{3}${\it  \normalsize Centro Tecnologico do Ex\'{e}rcito -- CTEx } \\
{\it \normalsize Av. das Americas 28705, 230020-470, Rio de Janeiro, RJ, Brasil}\\
$^{4}${\it  \normalsize Departamento de F\'{\i}sica, Universidade Federal Rural do Rio de Janeiro } \\
{\it \normalsize BR 465-07, 23851-180, Serop\'{e}dica, Rio de Janeiro, Brazil.}   }


\maketitle

\begin{abstract}
The Abelian Born-Infeld classical non-linear electrodynamic has been used to
investigate the electric and magnetostatic fields generated by a point-like
electrical charge at rest in an inertial frame. The results show a rich
internal structure for the charge. Analytical solutions have also been found.
Such findings have been interpreted in terms of vacuum polarization and
magnetic-like charges produced by the very high strengths of the electric
field considered. Apparently non-linearity is to be accounted for the
emergence of an anomalous magnetostatic field suggesting a possible connection
to that created by a magnetic dipole composed of two mognetic charges with
opposite signals. Consistently in situations where the Born-Infeld field
strength parameter is free to become infinite, Maxwell`s regime takes over,
the magnetic sector vanishes and the electric field assumes a Coulomb behavior
with no trace of a magnetic component. The connection to other monopole
solutions, like Dirac`s, t' Hooft`s or Poliakov`s types, are also discussed. Finally some speculative
remarks are presented in an attempt to explain such fields.

\end{abstract}





\footnote{This paper was submetted to International Journal of Theoretical Physics. }

\section{\label{sec:level1}INTRODUCTION}

\qquad This work investigates, under a classical approach and exploring the
non-linear properties of the Abelian B-I Theory, the configuration of the
fields generated by a single electric charge at rest. The main motivation for
this paper came from the question if a pure electic point-like charge at rest
in an inertial frame generates some kind of magnetostatic field. The challenge
resides on find classical solutions from B-I magnetic sector without
speculative additional assumptions. Only a few suitable restrictions have been
imposed. The results can be listed as following:

\qquad a) There is analytical and real stable magnetostatic dipole-like
solution generated by intense electric field strength;

\qquad b) It is possible to define a single magnetic charge in terms of the
electric point-like charge;

\qquad c) The findings is consistent concerning to the Maxwell linear theory.
In other words, it vanishes when the B-I parameter $b$ is free to become infinite.

\qquad d) The magnetic charge intensity calculated is close Dirac's prediction.

\qquad Born-Infeld (B-I) non-linear classical electrodynamics \cite{ref1,ref2}
represents an advanced theory to explain the structure and the finite energy
of the electron. It emerges in the more broad context of the M-Theory, where
the Superstrings Theories are enclosed. Recent revival of nonlinear
electrodynamics has been verified, mainly due to the fact that these theories
appear as effective theories at different levels of string/M-theory, in
particular in Dp-branes and super-symmetric extensions, and non-Abelian
generalizations. B-I Lagrangian describes the electromagnetic fields that live
on the world-volume of D-branes and T-duality gives direct evidence that it
governs the dynamics of the electromagnetic fields on
D-branes\cite{ref1.1,ref11}. B-I Lagrangian density is one of the general
non-derivative Lagrangians which depend only on the two algebric Maxwell
invariants. Among others its most attractive properties B-I Lagrangian is one
of the simplest non-polynomials that preserve gauge and Lorentz invariance,
the vacuum is characterized by $f_{\mu\nu}=0$ and the energy density is
positive. The field strength $f_{\mu\nu}$ is finite everywhere and is
characterized by its length $r_{o}$. B-I Theory is the only non linear
electrodynamic theory ensuring the absence of bi-refringence, this is, the
vacuum light speed is always $c$.

\qquad The organization of this paper is as follow: in Section 2 one accounts
for the exposition of the problem and the formulation of the main assumptions
and constraints. In Section 3 one gets the angular and radial differential
equations and solves then. Section 4 sets up the connection of the angular
moment and magnetic charge. Section 5 fixes the single magnetic charge and
Section 6 makes some final considerations.

\subsection{\label{sec:level2}CLASSICAL BORN-INFELD EQUATIONS IN MINKOVSKI
SPACE-TIME}

\qquad In these two sub-sections one exposes the standard Abelian B-I theory
embeded in flat space. The signature of the metric tensor and the first
assumption is established. The constitutive relations are suitable constructed
in order to make sure the integrability of the system.

\subsubsection{\label{sec:level3}THE ELECTRIC CHARGE AT REST}

\qquad The B-I non-linear electrodynamics action \cite{ref1} is defined, in
Minkovski space-time, as:%

\[
S=\int d^{4}xb^{2}\left[  1-\sqrt{-\det\left(  \eta_{\mu\nu}+\frac{f_{\mu\nu}%
}{b}\right)  }\right]
\]

\qquad The metric tensor $\eta_{\mu\nu}$ has signature (1,-1,-1,-1) and
$f_{\mu\nu}$ is the electromagnetic tensor. The parameter $b$, like the speed
of light in Einstein's relativity theory, is the maximum field strength
allowed by the B-I Theory and has a large value (about $10^{15}$ $esu$).
Setting its value to infinite leads to Maxwell`s linear electrodynamics. That
means that there is no limit to the field strength in Maxwell linear
electrodynamics. Enclosed in the action integral is the Born-Infeld Lagrangian
and evaluating its determinant yields:%

\[
L=b^{2}\left[  1-\sqrt{1-\frac{E^{2}-B^{2}}{b^{2}}-\left(  \frac
{\overrightarrow{E}\cdot\overrightarrow{B}}{b^{2}}\right)  ^{2}}\right]
\]

\qquad In addition the electric induction $\overrightarrow{D}$ and the
magnetic field $\overrightarrow{H}$, as derived from the canonical relation, are:%

\begin{equation}
\overrightarrow{D}=\frac{\partial L}{\partial\overrightarrow{E}}%
=\frac{\overrightarrow{E}+\left(  \frac{\overrightarrow{E}\cdot\overrightarrow
{B}}{b^{2}}\right)  \overrightarrow{B}}{\sqrt{1-\frac{E^{2}-B^{2}}{b^{2}%
}-\left(  \frac{\overrightarrow{E}\cdot\overrightarrow{B}}{b^{2}}\right)
^{2}}} \label{2.1.1}%
\end{equation}

\begin{equation}
\overrightarrow{H}=\frac{\partial L}{\partial\overrightarrow{B}}%
=\frac{\overrightarrow{B}-\left(  \frac{\overrightarrow{E}\cdot\overrightarrow
{B}}{b^{2}}\right)  \overrightarrow{E}}{\sqrt{1-\frac{E^{2}-B^{2}}{b^{2}%
}-\left(  \frac{\overrightarrow{E}\cdot\overrightarrow{B}}{b^{2}}\right)
^{2}}} \label{2.1.2}%
\end{equation}

\qquad The interaction with other charged particles is introduced by adding a
term $j_{\mu}A^{\mu}$ to the B-I Lagrangian. The equations of motion are the
standard Maxwell equations and the non-linearity is inserted in equations
(\ref{2.1.1}) and (\ref{2.1.2}). For a static point-like charge these
equations, on macroscopic level, are:%

\begin{align*}
\overrightarrow{\nabla}\cdot\overrightarrow{D}  &  =e\delta(\overrightarrow
{x})\qquad\overrightarrow{\nabla}\times\overrightarrow{E}=\overrightarrow{0}\\
\overrightarrow{\nabla}\cdot\overrightarrow{B}  &  =0\qquad\qquad
\overrightarrow{\nabla}\times\overrightarrow{H}=\overrightarrow{0}%
\end{align*}
and the solution for the electric induction $\overrightarrow{D}$, by taking
$\overrightarrow{B}=\overrightarrow{H}=\overrightarrow{0}$, is well known
\cite{ref1,ref2,ref3}. It is singular and identical to the Maxwell solution,
while the field $\overrightarrow{E}$ remains well defined at all points, even
at $%
r=0%
$. Thus:%

\begin{align}
\overrightarrow{D}  &  =\frac{e}{4\pi r^{2}}\widehat{r}\qquad and\qquad
\overrightarrow{E}=\frac{e}{\sqrt{r^{4}+r_{o}^{4}}}\widehat{r}\label{2.1.3}\\
r_{o}  &  =\sqrt{\frac{e}{4\pi b}}\qquad\widehat{r}=\overrightarrow
{r}/\left\vert \overrightarrow{r}\right\vert \nonumber
\end{align}

\qquad These are the necessary tools, for the present work, to describe that
electric charge in more broad way. No change will be done on electric sector
and it will be preserved by appropriate assumptions raised further.

\subsubsection{\label{sec:level4}THE NON ZERO MAGNETOSTATIC SECTOR}

\qquad What are the consequences of not setting $\overrightarrow{B}$ and
$\overrightarrow{H}$ equal to zero? What kind of fields could the theory
provide? If real and finite solutions do exist then what originates those
fields? In order to try to answer such questions one must consider each
component of the constitutive equations (\ref{2.1.1}) and (\ref{2.1.2}) at a
time. By assuming that the magnetic sector has only radial and polar
components that are functions of the radial distance $r$ from the point-like
charge, and the polar angle $\theta$, yields the following relations for
$\overrightarrow{B}$ and $\overrightarrow{H}$:%

\begin{align}
\overrightarrow{H}(r,\theta)  &  =H_{r}(r,\theta)\widehat{r}+H_{\theta
}(r,\theta)\widehat{\theta}\label{2.2.1}\\
\overrightarrow{B}(r,\theta)  &  =B_{r}(r,\theta)\widehat{r}+B_{\theta
}(r,\theta)\widehat{\theta}\nonumber
\end{align}

Where the subscripts refer to the components of the vector. So the problem is
axially symmetric. The introduction of $\varphi$ dependence will violate the
$\overrightarrow{\nabla}\times\overrightarrow{H}=\overrightarrow{0}$.

\qquad Then the constitutive relation (\ref{2.1.1}) and (\ref{2.1.2}) for each
component is written leading to an algebric non-linear system of equations
relating all electric and magnetic components.%

\begin{align}
E_{r}  &  =D_{r}\frac{R}{\left(  1+\frac{B_{r}^{2}}{b^{2}}\right)  }\qquad
E_{\theta}\approx0\\
B_{r}  &  =H_{r}\frac{R}{\left(  1-\frac{E_{r}^{2}}{b^{2}}\right)  }\qquad
B_{\theta}=H_{\theta}R
\end{align}

with%

\[
R=\sqrt{\left(  1+\frac{B_{r}^{2}}{b^{2}}\right)  \left(  1-\frac{E_{r}^{2}%
}{b^{2}}\right)  +\frac{B_{\theta}^{2}}{b^{2}}}%
\]

\qquad The magnetostatic field components ($B_{r}$ and $B_{\theta}$) are then
required to satisfy the following constraints $B_{r}\ll b$ and $B_{\theta}\ll
b$ at all points, while preserving the major feature of the theory, that is,
its nonlinearity. This is the first assumption introduced and is necessary in
order to ensure that the system remains integrable. The aforementioned set of
equations is then reduced to:%

\[
R=\sqrt{\left(  1+\frac{B_{r}^{2}}{b^{2}}\right)  \left(  1-\frac{E_{r}^{2}%
}{b^{2}}\right)  +\frac{B_{\theta}^{2}}{b^{2}}}\approx\sqrt{1-\frac{E_{r}%
^{2}(r)}{b^{2}}}%
\]

\begin{equation}
E_{r}(r)=\frac{D_{r}(r)}{\sqrt{1+\frac{D_{r}^{2}(r)}{b^{2}}}} \label{2.2.2}%
\end{equation}

\begin{equation}
B_{r}(r,\theta)=H_{r}(r,\theta)\sqrt{1+\frac{D_{r}^{2}(r)}{b^{2}}}
\label{2.2.2a}%
\end{equation}

\begin{equation}
B_{\theta}(r,\theta)=\frac{H_{\theta}(r,\theta)}{\sqrt{1+\frac{D_{r}^{2}%
(r)}{b^{2}}}} \label{2.2.2b}%
\end{equation}

\qquad That action keeps the B-I original behavior of the electric sector
while still preserving the connection between the magnetic and electric
sectors. At this point one concludes that, under the first assumption,
equations (\ref{2.2.2},\ref{2.2.2a},\ref{2.2.2b}) tell that the electric
sector shall not be affected by the induced magnetic sector. However each one
depends on $D_{r}(r)$, a subtle message from the electric to the magnetic sector.

\section{SOLUTION TO THE BORN-INFELD EQUATION}

\qquad This section is devoted to construct and solve the main differential
equation. One more basic assumption is necessary.in order to turn the system integrable.

\qquad The field $\overrightarrow{H}(r,\theta)$ is to satisfy $\overrightarrow
{\nabla}\times\overrightarrow{H}(r,\theta)=\overrightarrow{0}$ and the field
$\overrightarrow{B}(r,\theta)$ must be such that $\overrightarrow{\nabla}%
\cdot\overrightarrow{B}(r,\theta)=0$. So one has two PDE for the components of
the fields.%

\[
\frac{1}{r}\left[  \partial_{r}(rH_{\theta}(r,\theta))-\partial_{\theta}%
H_{r}(r,\theta)\right]  =0
\]

\[
\frac{1}{r^{2}}\partial_{r}[r^{2}B_{r}(r,\theta)]+\frac{1}{r\sin(\theta
)}\partial_{\theta}[\sin(\theta)B_{\theta}(r,\theta)]=0
\]

\qquad Now one assumes that the variables can be separable. This is the second
assumption of this paper. The magnetic field components can be written as:%

\begin{equation}
H_{r}(r,\theta)=h_{r}(r)G(\theta)\qquad H_{\theta}(r,\theta)=h_{\theta
}(r)J(\theta)
\end{equation}

The magnetic induction components are then determined by the magnetic field
and become:%

\begin{equation}
B_{r}(r,\theta)=b_{r}(r)G(\theta)=h_{r}(r)\sqrt{1+\frac{D_{r}^{2}(r)}{b^{2}}%
}G(\theta) \label{2.2.3}%
\end{equation}

\begin{equation}
B_{\theta}(r,\theta)=b_{\theta}(r)J(\theta)=\frac{h_{\theta}(r)}{\sqrt
{1+\frac{D_{r}^{2}(r)}{b^{2}}}}J(\theta) \label{2.2.4}%
\end{equation}

Where $h_{r}(r)$, $h_{\theta}(r)$, $G(\theta)$ and $J(\theta)$ are unknown
functions to be determined and $b_{r}(r)=h_{r}(r)\left(  1+\frac{D_{r}^{2}%
(r)}{b^{2}}\right)  ^{1/2}$ as well $b_{\theta}(r)=h_{\theta}(r)\left(
1+\frac{D_{r}^{2}(r)}{b^{2}}\right)  ^{-1/2}$.

\qquad Then substituting $H_{r}(r,\theta)$ and $H_{\theta}(r,\theta)$ on
$\overrightarrow{\nabla}\times\overrightarrow{H}(r,\theta)=\overrightarrow{0}%
$, yields a set of two differential equations where $\lambda$ is a constant.%

\begin{equation}
\frac{1}{h_{r}(r)}\frac{d[rh_{\theta}(r)]}{dr}=\frac{1}{J(\theta)}%
\frac{dG(\theta)}{d\theta}=\lambda\label{4.1}%
\end{equation}

\qquad Likewise, the substitution of $B_{r}(r,\theta)$ and $B_{\theta
}(r,\theta)$ in $\overrightarrow{\nabla}\cdot\overrightarrow{B}(r,\theta)=0$
leads to:%

\begin{equation}
\frac{1}{rb_{\theta}(r)}\frac{d}{dr}\left[  r^{2}b_{r}(r)\right]  =-\frac
{1}{\sin(\theta)G(\theta)}\frac{d(\sin(\theta)J(\theta))}{d\theta}%
=\varsigma\label{4.2}%
\end{equation}

Where $\varsigma$ is another constant. That can be resumed in \ a system of
four differential equations further.%

\begin{equation}
\frac{d[rh_{\theta}(r)]}{dr}=\lambda h_{r}(r) \label{4.2.2}%
\end{equation}

\begin{equation}
\frac{dG(\theta)}{d\theta}=\lambda J(\theta) \label{4.2.3}%
\end{equation}

\begin{equation}
\frac{d}{dr}\left[  r^{2}b_{r}(r)\right]  =\varsigma rb_{\theta}(r)
\label{4.2.4}%
\end{equation}

\begin{equation}
\frac{d(\sin(\theta)J(\theta))}{d\theta}=-\varsigma\sin(\theta)G(\theta)
\label{4.2.5}%
\end{equation}

\qquad Considering the angular equations (\ref{4.2.3}) and (\ref{4.2.5}) one
arrives on a second order differential equation for $G(\theta)$.%

\begin{equation}
\frac{d}{d\theta}\left\{  \sin(\theta)\frac{d(G(\theta))}{d\theta}\right\}
+\lambda\varsigma\sin(\theta)G(\theta)=0 \label{4.3}%
\end{equation}
which has the general angular solution given in terms of the Legendre
functions of the first kind $P_{n}(\cos(\theta))$ and Legendre function of
second kind $Q_{n}(\cos(\theta))$. The second has has the undesirable feature
of not be always real.%

\[
G_{n}(\theta)=C_{1}P_{n}(\cos(\theta))\qquad n=\frac{\sqrt{1+4\lambda
\varsigma}}{2}-\frac{1}{2}%
\]

\qquad Doing the same action on radial equations (\ref{4.2.2}) and
(\ref{4.2.4}), with the support of the relation between $b_{i}(r)$ and
$h_{i}(r)$, the result is a differential equation for $\Psi(r)=rh_{\theta}(r)$.%

\begin{equation}
\frac{d}{dr}\left\{  \sqrt{r^{4}+r_{o}^{4}}\frac{d\Psi\left(  r\right)  }%
{dr}\right\}  -\frac{\lambda\varsigma\Psi(r)r^{2}}{\sqrt{r^{4}+r_{o}^{4}}}=0
\label{4.3.1}%
\end{equation}
The general solution is the associated Legendre functions of the first and
second kind $P_{n}^{1/4}(z)$ and $Q_{n}^{1/4}(z)$, with $z=\sqrt{\left(
\frac{r}{r_{o}}\right)  ^{4}+1}$ and $n$ exactly the same for the angular
sector. Once again the second one takes complex values and the solution can be
written as:%

\begin{equation}
\Psi_{n}(r)=\overline{m}\sqrt{r}P_{n}^{\frac{1}{4}}\left(  z\right)  \qquad
z=\sqrt{\left(  \frac{r}{r_{o}}\right)  ^{4}+1} \label{4.3.2}%
\end{equation}

\qquad Acceptable solutions for (\ref{4.3}) require that $n$ must be a natural
number, otherwise the field lines will not be closed. As result $\lambda
\varsigma$ assumes special integers $(0,2,6,12,...)$. Looking for solutions
(\ref{4.3.2}) the component $h_{\theta}(r)$ is calculated and from
(\ref{2.2.4}) the component $b_{\theta}(r)$ are obtained. However, among this
infinite set only two solutions will give physical meaning. For $\lambda
\varsigma\geqq6$ the component $h_{\theta}(r)$ grows as the radial distance
become large.\ For $\lambda\varsigma=0$ one gets no angular dependence,
$P_{0}(\cos(\theta))=1$, and the field will be entirely radial like a monopole
field. In this situation there is no angular moment and no way to express the
magnetic charge in terms of electric charge. On the other side the first
assumption is violated when $r\approx0$ because equation (\ref{4.2.4}) tells
that $b_{r}(r)$ grows without limit. Taking $\lambda\varsigma=2$ the situation
is different, $G_{1}(\theta)=C_{1}\cos(\theta)$, and equation (\ref{4.2.3})
gives the second angular function such that, $J_{1}(\theta)=-C_{1}\sin
(\theta)/\lambda$.\ The solution (\ref{4.3.2}) will be\cite{ref10}:%

\[
\Psi(r)=\overline{m}\sqrt{r}P_{\frac{1}{4}}^{\frac{1}{4}}\left(  z\right)
\]
and the constant $\overline{m}$ will be closely related to the magnetic dipole
moment for suitable choice of the constants $\lambda$ and $\varsigma$. It is
then possible to represent the magnetostatic field polar component,
$b_{\theta}(r)$, like this:

\begin{center}
$b_{\theta}(r)=\frac{h_{\theta}(r)}{\sqrt{1+\frac{D_{r}^{2}(r)}{b^{2}}}}%
=\frac{r}{\sqrt{r^{4}+r_{o}^{4}}}\Psi(r)$
\end{center}

\qquad The differential equation (\ref{4.2.4}) allows $b_{r}(r)$ calculation.

\qquad Each component, polar and radial, is finite everywhere and the Figure 1
shows the field strength of the total magnetostatic field, in arbitrary units,
as a function of the radial distance of the electric point-like charge.

\begin{center}

\begin{center}
\begin{figure}[h]

\vspace{0.3cm}
{\par\centering
\resizebox*{0.60\textwidth}{!}{\rotatebox{-90}{\includegraphics{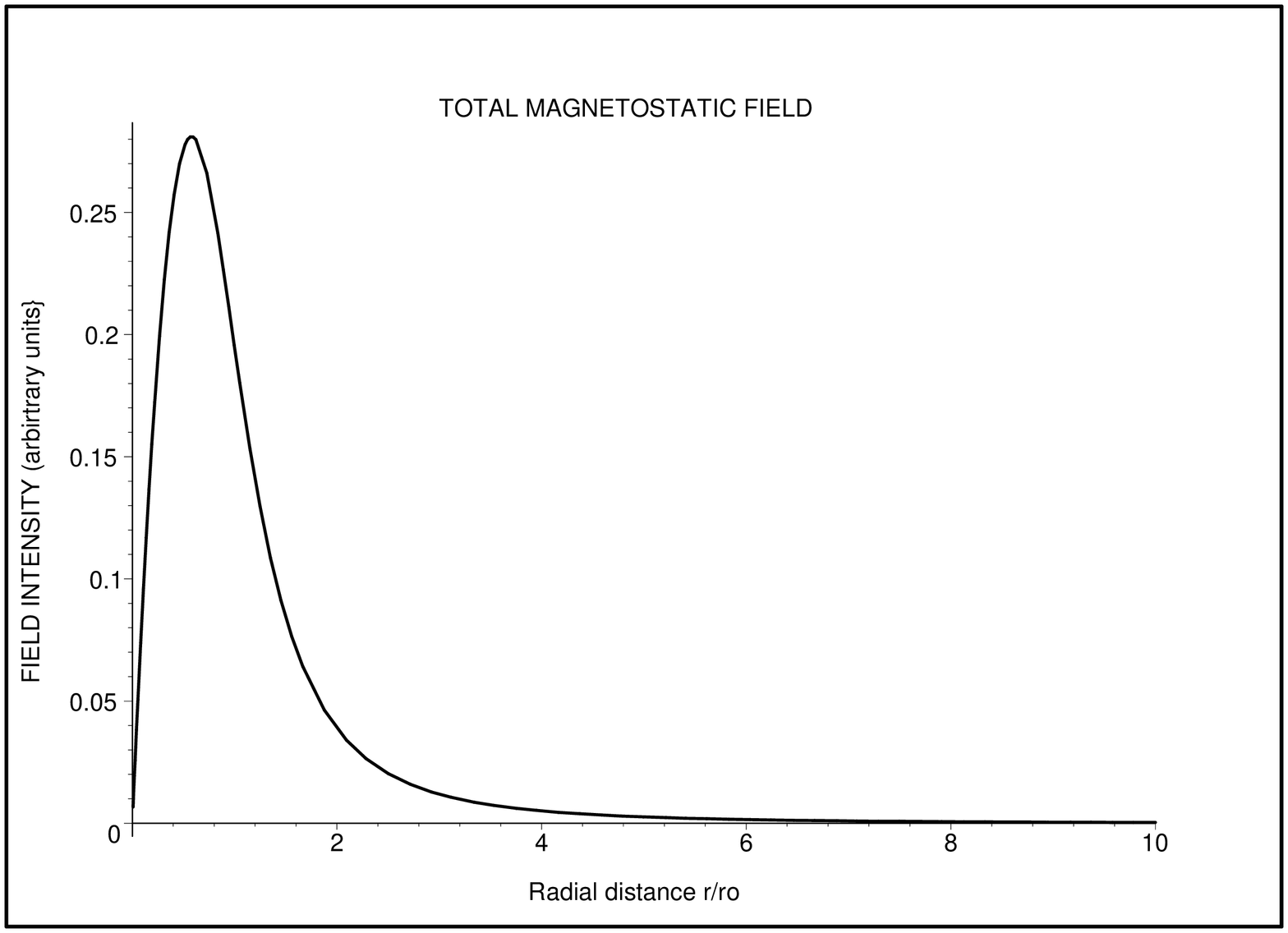}}}
\par}
{\par\ \par}
{\par\centering Figure 1\par}
\vspace{1cm}
\end{figure}

\end{center}

\end{center}

Each one has a maximum near the $r_{o}$ and vanishes as $r\rightarrow0$ or
$r\rightarrow\infty$. An accurate mathematical analysis shows that the
asymptotic behaviour is the well known $r^{-3}$, for $r\gg r_{o}$, indicating
the connection of $\overline{m}$ with some magnetic dipole moment, and
proportional to $r$ for $r\approx0$, indicating some induced magnetic charge
distribution. Also, the nonlinearity effect can be viewed near $r_{o}$ where
polar and radial field intensity change its relative magnitude. Out of that
range the components recover their linearity. To see this one takes
$h_{\theta}(r)$ as:%

\[
h_{\theta}(r)=\frac{\overline{m}r_{o}^{2}f(z)}{r^{3}}\qquad f(z)=z^{2}\left[
\sqrt{z}P_{\frac{1}{4}}^{\frac{1}{4}}\left(  \sqrt{1+z^{4}}\right)
-kz\right]
\]

The function $f(z)$, called here as "$form\ fuction$" can be showed in Figure
2. The asymptotic behavior is constant indicating that closely connection of
$\overline{m}$ with the magnetic dipole moment.%

\begin{center}
\begin{figure}[h]
\vspace{0.3cm}
{\par\centering
\resizebox*{0.60\textwidth}{!}{\rotatebox{-90}{\includegraphics{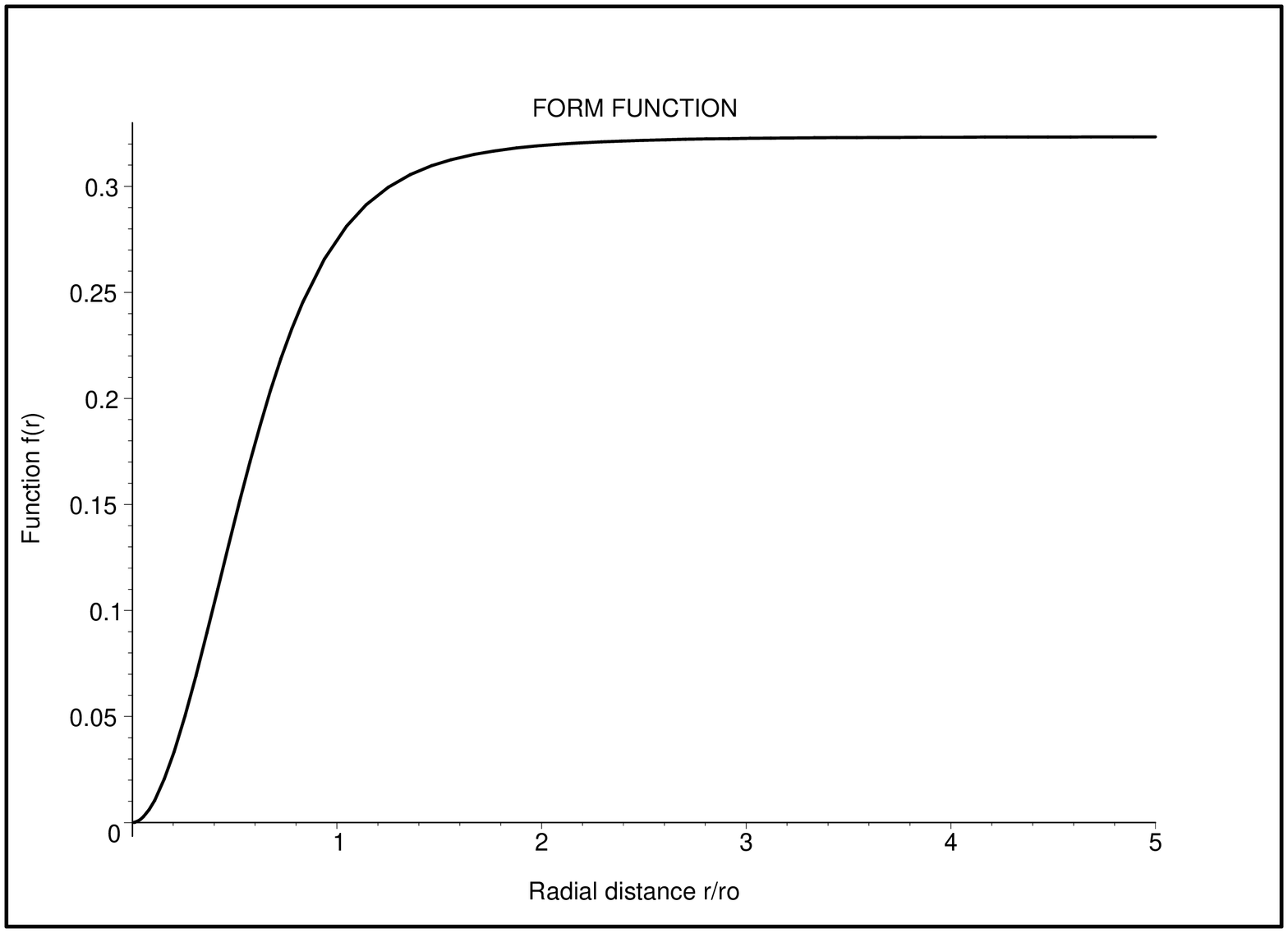}}}
\par}
{\par\ \par}
{\par\centering Figure 2\par}
\vspace{1cm}
\end{figure}
\end{center}

\qquad Tangent field lines can be evaluated equating the ratio of field
components to the slop of the curve in polar coodinates.%

\begin{equation}
\frac{B_{r}(r,\theta)}{B_{\theta}(r,\theta)}=\frac{dr}{rd\theta}%
\end{equation}

\qquad Integrating the last equation for two different assymptotic distances
from the electric charge one arrives to the following equations in polar coordinates:

\begin{center}
$r=k_{1}\sin^{2}(\theta)\qquad r\gg r_{o}$

$r=k_{2}\sin^{2/3}(\theta)\qquad r\approx0$
\end{center}

\qquad Figure 3 shows the tangent closed lines for the total magnetic
induction. It's a tipical magnetic dipole pattern. The constants $k_{1}$ and
$k_{2}$ are set up for different range of radial distance. For\ small loop the
entire curve is near correct. The other has some restrictions for polar angles
near north and southern pole.

\qquad The interpretation carried out was as two magnetic charge of oposite
signal distributed over each hemisphere. The vacuum so polarized by the
intense electric field behaves like a material surrounding the electric
charge. These induced magnetic field can produce angular moment when coupling
with the electric field.

\begin{center}
\begin{figure}[h]
\vspace{0.3cm}
{\par\centering
\resizebox*{0.60\textwidth}{!}{\rotatebox{-90}{\includegraphics{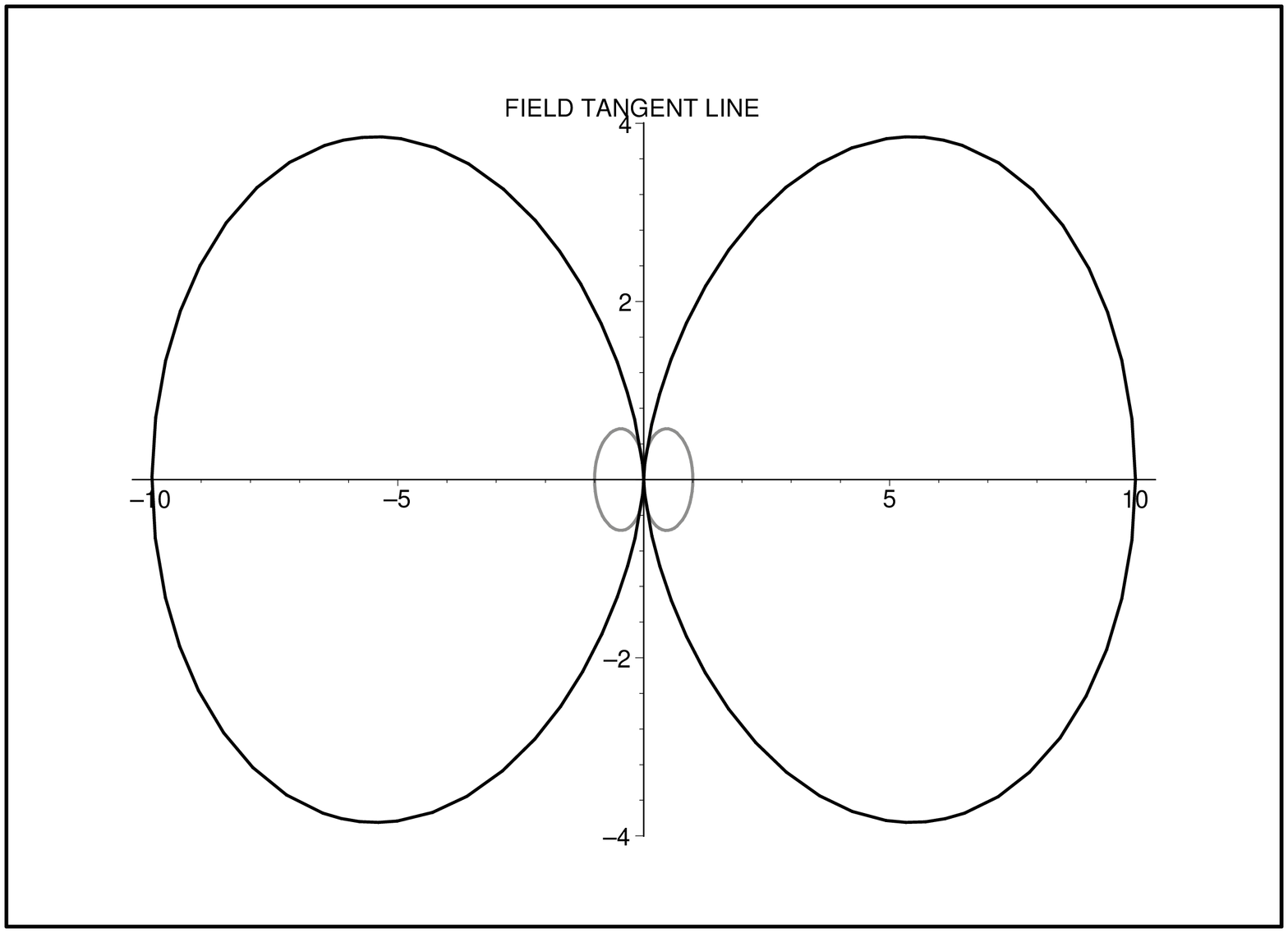}}}
\par}
{\par\ \par}
{\par\centering Figure 3\par}
\vspace{1cm}
\end{figure}
\end{center}

\qquad The interpretation carried out was as two magnetic charge of oposite
signal distributed around the poles. The vacuum so polarized by the intense
electric field behaves like a material surrounding the electric charge. These
induced magnetic field can produce angular moment when coupling with the
electric field.

\section{THE ANGULAR MOMENT AND THE CONNECTION WITH THE MAGNETIC CHARGE}

\qquad In this section the proposed induced magnetic charge, interpreted in
last section, is written as a combination of fundamental parameters like the
electric charge $e$ and the B-I maximum field strength $b$. In order to
understand what that solution means, the angular moment of those fields must
be calculated due to its great importance \cite{ref4,ref5,ref6}. From the
definition $\overrightarrow{L}$ can be evaluated and its interpretation
follows immediately.%

\begin{equation}
\overrightarrow{L}=\int\overrightarrow{x}\times\left(  \overrightarrow
{D}\times\overrightarrow{B}\right)  d^{3}\overrightarrow{x} \label{5.1}%
\end{equation}

\qquad Using the properties of symmetry of the above integral yields the
angular moment, that is completely aligned to the polar $z$ axis.%

\begin{equation}
L_{z}=2\pi\alpha\mu_{0}\overline{m}r_{o}e \label{5.2}%
\end{equation}

\[
\alpha=%
{\displaystyle\int\limits_{0}^{\infty}}
dz\frac{f(z)}{\sqrt{1+z^{4}}}\approx0.4003
\]
and $\mu_{0}$ is the magnetic permeability

\qquad The result (\ref{5.2}) suggests that an interpretation be given in
terms of a magnetic charge. It is known that the classical angular moment
\cite{ref6} for a system consisting of an electric monopole and a magnetic
monopole is $ge$, the parameter $g$ being the strength of the magnetic charge.
If both results are compared, it is then possible to explain such magnetic
charge as the result of a non-linearity effect or of the vacuum polarization
of the B-I theory.%

\begin{equation}
g_{eff}\rightarrow2\pi\alpha\mu_{0}\overline{m}r_{o} \label{5.4}%
\end{equation}

\qquad That is consistent with the rigorously linear behavior found when the
maximum field strength is allowed to become infinite. In that case the
contribution to the magnetic sector vanishes and that of the magnetic charge
completely disappear. Based on the findings presented, a proposal for a model
of the structure of the electric charge could consist of an electric monopole,
of strength $e$, and two distributed magnetic charge $g$, of opposite signs,
in each hemisphere, yielding a total null magnetic charge when seen at a great
distance, and null net divergence, like a magnet with its closed field lines.

\section{FIELD AS A SOURCE OF FIELD}

\qquad Take advantage of the result of the previous section and of the
separation of variables, one speculates, in this section, on the possibility
of get a single magnetic charge. There is another way to interpret the result
of the previous section. Looking for the divergence of the vector
$\overrightarrow{B}(r,\theta)$ and removing the angular solution covering it,
one gets an equation with $-\frac{2}{r}b_{\theta}(r)$ as a source of the pure
radial field. That claims for a divergence of some radial field,
$\overrightarrow{\nabla}_{r}\cdot\overrightarrow{B}_{r}$. So one has the
transverse field component as a source of a pure radial field given by a
spread magnetic density $\rho_{m}(r)$.%

\begin{equation}
\frac{1}{r^{2}}\frac{d[r^{2}b_{r}(r)]}{dr}+\frac{2}{r}b_{\theta}(r)=0
\label{6.1}%
\end{equation}

\begin{equation}
\overrightarrow{\nabla}\cdot\left[  b_{r}(r)\widehat{r}\right]  =\rho
_{m}(r)\qquad\rho_{m}(r)=-\frac{2}{r}b_{\theta}(r) \label{6.2}%
\end{equation}

\qquad The total magnetic charge strength , $g$, is obtained integrating
$\rho_{m}(r)$ over the entire space.%

\begin{equation}
g^{\prime}=\int d^{3}\overrightarrow{x}\left[  \frac{2}{r}b_{\theta
}(r)\right]  =8\pi\alpha\mu_{0}\overline{m}r_{o} \label{6.3}%
\end{equation}

\qquad It differs from (\ref{5.4}) only in magnitude, it is four times bigger
than $g_{eff}$. Our interpretation is that the symmetry cancels all three
directions contributions to the angular moment. When the symmetry is broken
this cancellation desapairs. Extracting it from the $z$ angular moment
component one looks only for a fraction of the total. The remained is hidden
by the symmetric components cancellation on interacting with the electric
charge to produce the angular moment. Moreover the shadowing between both
magnetic charges reduce the effective individual magnetic charge ntensity
given $g^{\prime}>g_{eff}$.

\section{NUMERICAL CALCULATIONS}

\qquad This section plays the calculation of the magnetic charge in SI units.
For this task it is necessary to recover the SI units of the objects in
angular momentum. The constant $\overline{m}r_{o}^{2}\alpha$ can be given
roughly as $er_{o}\alpha c$, where $e$ is the elementary charge, $r_{o}$ is
the B-I radius parameter, $c$ is the speed of ligth and $\alpha$ is that
integral involving the Form Function defined previously. We interpret $\alpha
c$ as the velocity of the spining charge $e$ around the circle of radius
$r_{o}$.

\qquad Inserting all in (\ref{5.4}) we calculated, with $\mu_{0}=4\pi
\times10^{-7}n/A^{2}$, the intensity of magnetic charge in consideration,
about $60e$. This is less than Dirac quantum prevision that is $68.5\ e$. A
more accurate result only a quantum version of that theory could give us.

\section{FINAL CONSIDERATIONS}

\qquad We showed that Classical Abelian Born-Infeld electrodynamics can
predicts the existence of real and well-behaved magnetostatic fields solutions
associated with electric charges at rest. Definitely it is a non-linear effect
simply ruled out by Maxwell's electrodynamics. Although Born-Infeld non-linear
electrodynamics has not yet been experimentally confirmed, it was originally
conceived to describe the electron properties based only on the structure of
the field. The findings of this work apparently suggest that the fields,
rather than resembling Dirac's or t'Hooft's magnetic monopoles
\cite{ref8,ref9}, exhibit properties similar to those produced by a magnet as
considered from the macroscopic point of view although its more complex
structure is only seen at the microscopic level. In addition they indicate
that the magnetostatic solutions ensued from the breaking of the radial
symmetry. That was necessary in order to see the dependence of the magnetic
charge on basic parameters such the maximum field strenght $b$ and the
electric charge $e$. Such results seem to suggest the existence of vacuum
excitation effects caused by the intense electric field strength around the
electric charge. In addition by singling out the angular dependence, the
current approach allows the investigation of a pure magnetic radial field
generated by a spatially distributed magnetic charge derived from an electric
charge. It must be stressed that quantum effects prevail over the classical
description when distances in the order of $10^{-15}m$ are considered and the
Compton wavelength of the electron, $\hbar/m_{e}c$, is about $10^{-12}m$.
Hence $r_{o}$ still lies well within the limits of the classical validity, and
a quantum Born-Infeld Theory still remains to be developed.

\qquad One of the major motivations for the use of the Born-Infeld
electromagnetic field theory is to overcome the infinity problem associated
with a point-like charge source as in Maxwell`s theory. Born's original theory
may currently be explained as an attempt to find classical solutions to
represent electrically charged states produced by sources that have finite
self-energy. The present work proposes an extension to that theory by also
seeking magnetically stable solutions derived for purely electrical charges.
As expected, all additional anomalous magnetic terms vanish when Maxwell's
regime is restored by allowing the maximum field strength to become infinite.

\bigskip \noindent{\bf Acknowledgments}\\
S.O. Vellozo wish to thank CBPF and CTEx by the support. L.P.G. De Assis is grateful to FAPERJ-Rio de Janeiro for his post-doctoral fellowship.

\end{document}